\documentclass[10pt]{article}
\usepackage{graphics}
\begin{document}

\title{\bf Stereography using a single lens}
\author{}
\date{}

\maketitle
\vspace{-1.8cm}

\begin{center}
\begin{tabular}[t]{c@{\extracolsep{2em}}c} 
  \bf    Kshitija Deshpande \footnotemark[1]    &\bf    Arvind Paranjapye \footnotemark[2]    \\
  \small\tt kbdeshpande@gmail.com & \small\tt arp@iucaa.ernet.in\\
\end{tabular}
\smallskip

\begin{small}
\begin{tabular}{rl}
\footnotemark[1] & \parbox[t]{13cm}{College of Engineering, Pune 411 005}\\
\footnotemark[2] & \parbox[t]{13cm}{Inter University Center for Astronomy and Astrophysics, Pune 411 007}\\
\end{tabular}
\end{small}
\bigskip
\end{center}

\begin{abstract}

In this paper we have put forth an innovative method of obtaining a stereographic image on a single frame using a single lens. This method has been verified experimentally. A preliminary prototype of the same is  built with an optimized use of the material available in the laboratory. The prospective applications of this technique are also explored in brief. This method once commercialized, will reduce the expenses incurred in the stereo videography. 

We also propose a simplified method of obtaining anaglyph.


\end{abstract}

\section{Introduction} 

Stereoscopic imaging strengthens the perception of dimensionality through the creation of two dimensional drawings, images or video on two adjacent frames that when viewed by both eyes appear to exhibit a feeling of depth. This seemingly simple phenomenon of depth perception is a product of a complicated set of interactions between our eyes and our brains that is still not entirely understood. Our eyes are spaced about $6\ cm$ $(2.5\ in)$ apart, which causes each eye to receive a slightly different image. The brain fuses these two images into a single 3D image, enabling us to perceive depth. This way of seeing is called binocular vision or stereoscopic vision or stereopsis.

3D films generally use two cameras or one camera with two lenses. The centers of the lenses are spaced $6.25\ cm\ to\ 6.75\ cm$ apart to replicate the displacement between a viewer's left and right eye. Each lens records a slightly different view, corresponding to the different view each eye sees in normal vision.

A camera with two lenses is usually used for stereography \cite{e}. There are techniques where stereographs are created by shifting the camera positions. Similar techniques are also used in 3D video rendering \cite {a}. We have used the basic principle of reflection to obtain two images of the same object but at different angles on a single frame.
 
In the following sections, we discuss the implementation of the prototype for stereoimaging with only one lens its results and its possible uses. We also propose a model for anaglyph imaging.

\section{More on 3D images:}
Three-dimensional images are of following types: 

{\bf Stereographs:} 
 	One-Way to impart the illusion of depth in a photograph is to create a stereograph \cite{d}, which is a combination of two photographs of the same scene, taken from slightly different angles. These are viewed through stereoscopes \cite{j}.

{\bf Anaglyphs:} 
 	Anaglyphs use colored filters to create the illusion of depth in motion pictures, photographs, and illustrations. An anaglyph consists of two slightly different views of the same scene printed in two different colors, generally red and blue, then superimposed onto one another. To appear 3D, anaglyph must be viewed through special glasses fitted with corresponding red and blue filters. The brain receives two separate images and fuses them to give us a perception of depth. Anaglyphs are used in comics, cartoons and for some scientific purposes. Anaglyph imaging was used by the Viking space probes to take thousands of anaglyphic photos of the surface of the mars, between 1975 and 1982. This provided astronomers with three dimensional panoramic views and helped to have an idea about the features of the surface of mars.

{\bf Polarized 3D films:}
Polarizing properties are also used for creating 3D images \cite{k}.

{\bf Autostereograms:}
An autostereogram is a stereoscopic image that does not require special viewing devices, such as a stereoscope or a 3D glasses. Among the most popular types of autostereograms are: Lenticular 3D images, Holograms, Computer Generated Single Image Random Dot Stereograms (SIRDS) \cite{b}.

\section{\bf Prototype design outline:}

While implementing this novel idea, we decided to avoid any sophistication. Using the following apparatus, we designed the prototype to evaluate the feasibility of our method for obtaining stereo video and/or stereo images.

{\bf Apparatus:}

\indent  Web camera with a computer or laptop. Two front coated silvered mirrors $(5\ cm\ \times\ 6\ cm)$. Two front coated silvered mirrors of twice the size of previous mirrors $(10\ cm\ \times\ 12\ cm)$. Blocks for holding mirrors and the web camera. A wooden board for arrangement of the apparatus.

{\bf Arrangement and working:}

\begin{figure}[h]
  \begin{center}
    \caption{The setup of prototype for Stereo videography\label{fig:stereogram}}
    \scalebox{0.75}{
      \includegraphics{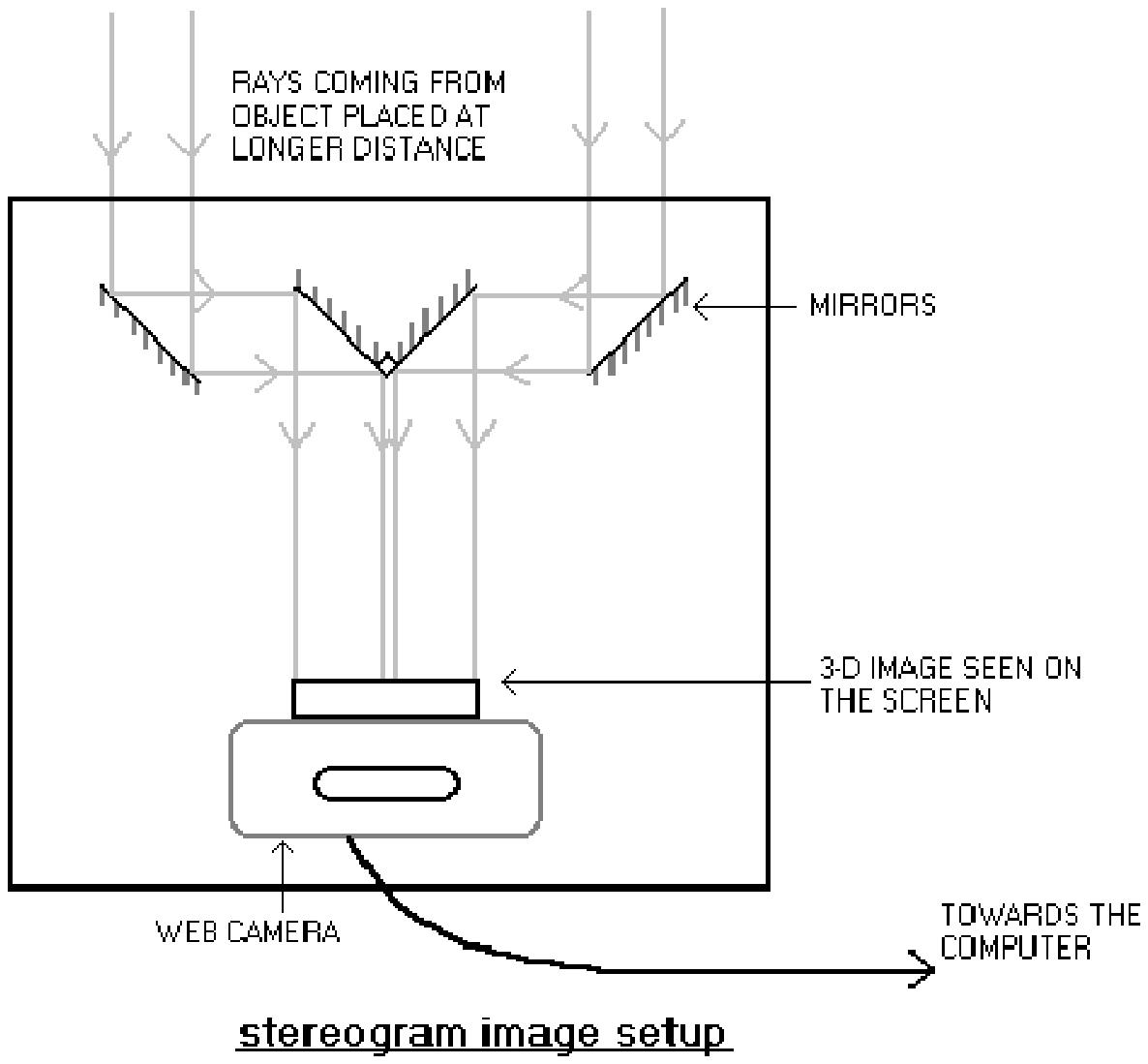}
    }
  \end{center}
 
\end{figure}

\indent  As shown in the Figure~\ref{fig:stereogram}, the two smaller mirrors $(5\ cm\ \times\ 6\ cm)$ are placed with the reflecting surfaces making an angle of $90 ^ {\circ}$ and the corner of their combination is kept so as to face the web camera. The other two mirrors $(10\ cm\ \times\ 12\ cm)$ are placed parallel to these mirrors approximately $25\ cm$ away such that they face the object. 

\indent With this arrangement described above, the rays from the object get reflected, so as to give two exactly identical images one besides the other, on the same frame on the screen of the computer where the web camera is connected. The 3D effect can be perceived by observing the left image with left eye and right image with right eye. Practically speaking, to see the 3D image, focus your eyes to infinity. Then look at the frame on the computer screen with two images side by side and moving back and forth till the two images merge in each other. At this moment, a 3D effect can be perceived. This can be achieved by squinting the eyes as well. 

\indent There are viewing softwares and stereoscopes available in market \cite{h} to perceive the stereographs without squinting the eyes.

\begin{figure}[p]
  \begin{center}
   \caption{\bf Stereographs- results in the form of clip from the prototype\label{fig:stereoclip}}
    \scalebox{0.75}{
      \includegraphics{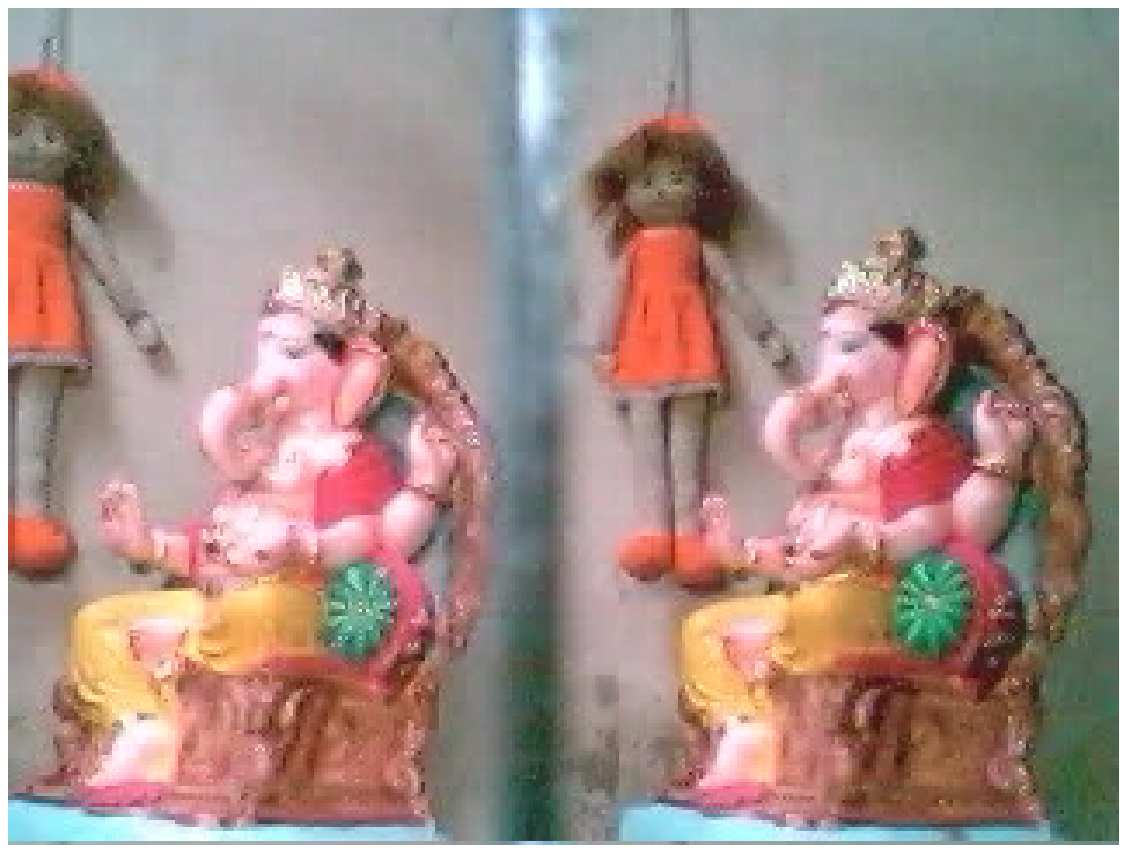}
    }
  \end{center}
 
   \begin{center}
    \scalebox{0.75}{
      \includegraphics{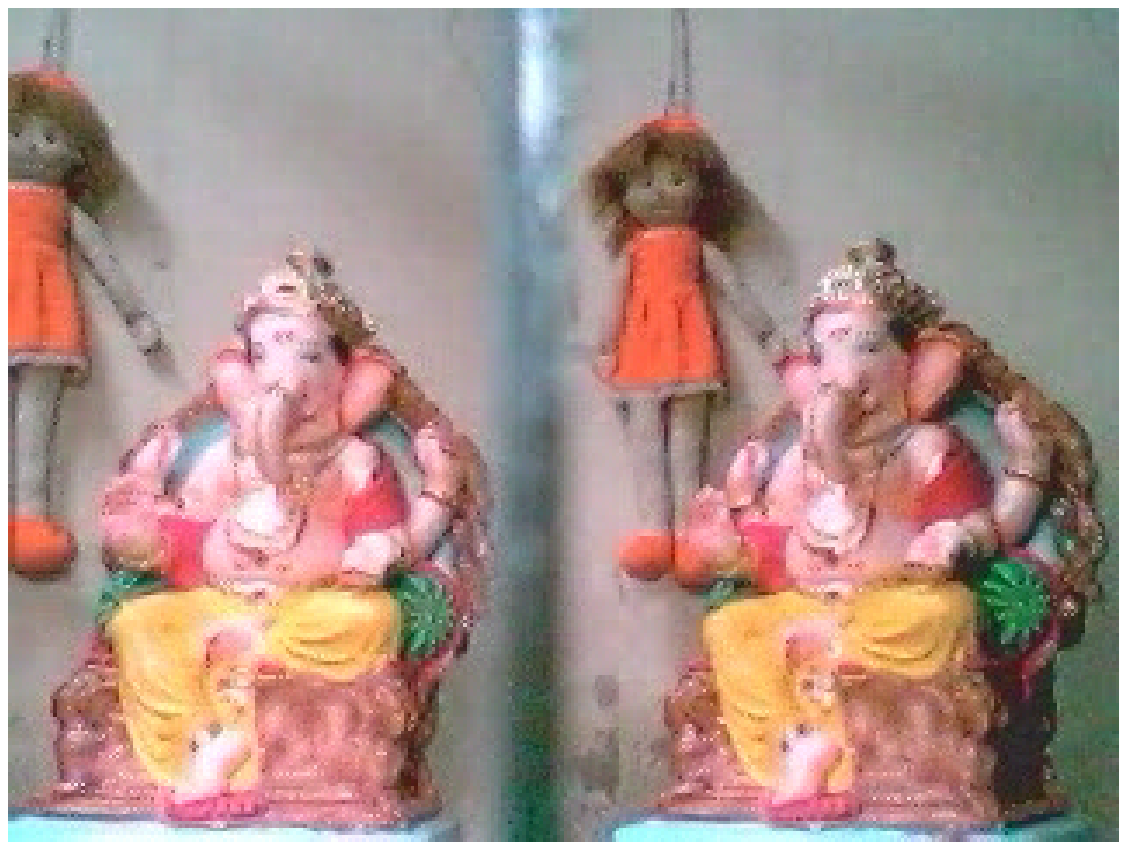}
    }
  \end{center}

\end{figure}

Figure~\ref{fig:stereoclip} shows the clippings of the stereographic movie taken with this prototype. Two slightly different images of the same object can be seen side by side. A slight difference is introduced because of the tilting (angular change) of one of the larger mirrors placed parallel to the pair of the smaller mirrors in Figure~\ref{fig:stereogram}. This is a stereograph, which when viewed with the aid of stereoscopes \cite{h} or without any aid of instruments, a perception of depth is observed in the merged image. The sequence of the images shown in Figure~\ref{fig:stereoclip} is an evidence of successful capture of a 3D movie. 

\section{\bf Innovative component in the design:} 

  The use of single lens eliminates the complicated procedure of aligning the two images for a stereograph when 3D images are taken using two independent optical systems. This is the most important innovative component of the design. The instruments currently in use for stereoscopic or anaglyphic imaging consist of two cameras kept at different positions to capture the same object at different viewing angles. There are two separate films with the image of the same object. Sometimes a specially made expensive stereoscopic camera with two embedded lenses is used. A complex mechanism is required for these types of cameras to combine the two images. For such type of imaging, beam splitters, projectors etc. are also needed. 

  Stereography performed using a single lens needs the camera position to be slightly shifted \cite{f} after capturing a frame. This drawback of the existing technique of stereography with single lens is removed in this idea.
  
  The method proposed here is based on the basic principle of reflection. The inherent simplicity of the design and the use of inexpensive available material has made this design ingeniously peculiar.

\section{\bf Future Scope}

  Picture quality and clarity can be improved by using high quality Charged Coupled Device (CCD) camera. Movable mirror system can be designed to have parallax at calculated distances. A portable and compact model for smaller CCD cameras namely the one on cellphone can be designed.~\\
\indent  The same equipment after some change can be used for viewing the 3D video of distant objects. It may be used for inter-orbital survey, by astronauts so as to detect old satellite debris.The technique could also help third umpires in the sports to remove an element of doubt. Thus with an improvement in the quality, a cheaper, ready to sell product can be produced.

\indent A better, compatible viewing instrument like Helmet Mounted Displays (HMD) \cite{c} can be devised to reduce the strain on eyes while observing stereographs. 

\section{\bf Possible Uses}

  The method of stereoimaging using only one lens can find uses in medical field for example, in microsurgery,like 3D on-screen microsurgery system (TOMS), which is a procedure that projects a magnified 3D image through the video microscope, on one or more TV monitors. This reduces the strain on eyes and also improves views of the operating field. It can also be used for educational purposes like study of live surgery for doctors and students. More advanced versions would be useful in 3D endoscopy and Stereo microscopes.

\indent  In various research fields including under water archeology, land archeology, ornithology and entomology, this technique of stereography can be used.

\indent  Stereography is also utilized in topographic survey where mapping is done using aerial photography. Two photographs when studied together in a stereoscope or with 3D display technologies, will give them an appearance of 3D and permit mapping height of terrain and buildings. Here for the aerial photography a robust version of this model could be used. The advent of DTI 2D/3D display technology enables us to get an actual feel of the terrains and buildings \cite{i}.

\indent  The technique can be used for the entertainment, travel \cite{g}, advertising, product marketing industries. The other possible applications of this innovation are in tele-robotics, in unmanned systems in defense, in tele-presence and Stereo video conferencing.

\section{\bf Proposed model for anaglyph} 

In the single lens stereograph, only half of the frame from an imaging device is used. Using partially transmitting and reflecting $(50-50\%)$ optical flat with two filters (blue and red) as shown in the figure below, full frame 3D anaglyphs can be obtained.

Though our design uses only half of the frame as against anaglyph design proposed here, it gives us 3D color images (still or moving) whereas anaglyphs are black and white images.

\begin{figure}[h]
  \begin{center}
    \caption{The proposed setup for Anaglyph imaging \label{fig:anaglyph}}
    \scalebox{0.75}{
      \includegraphics{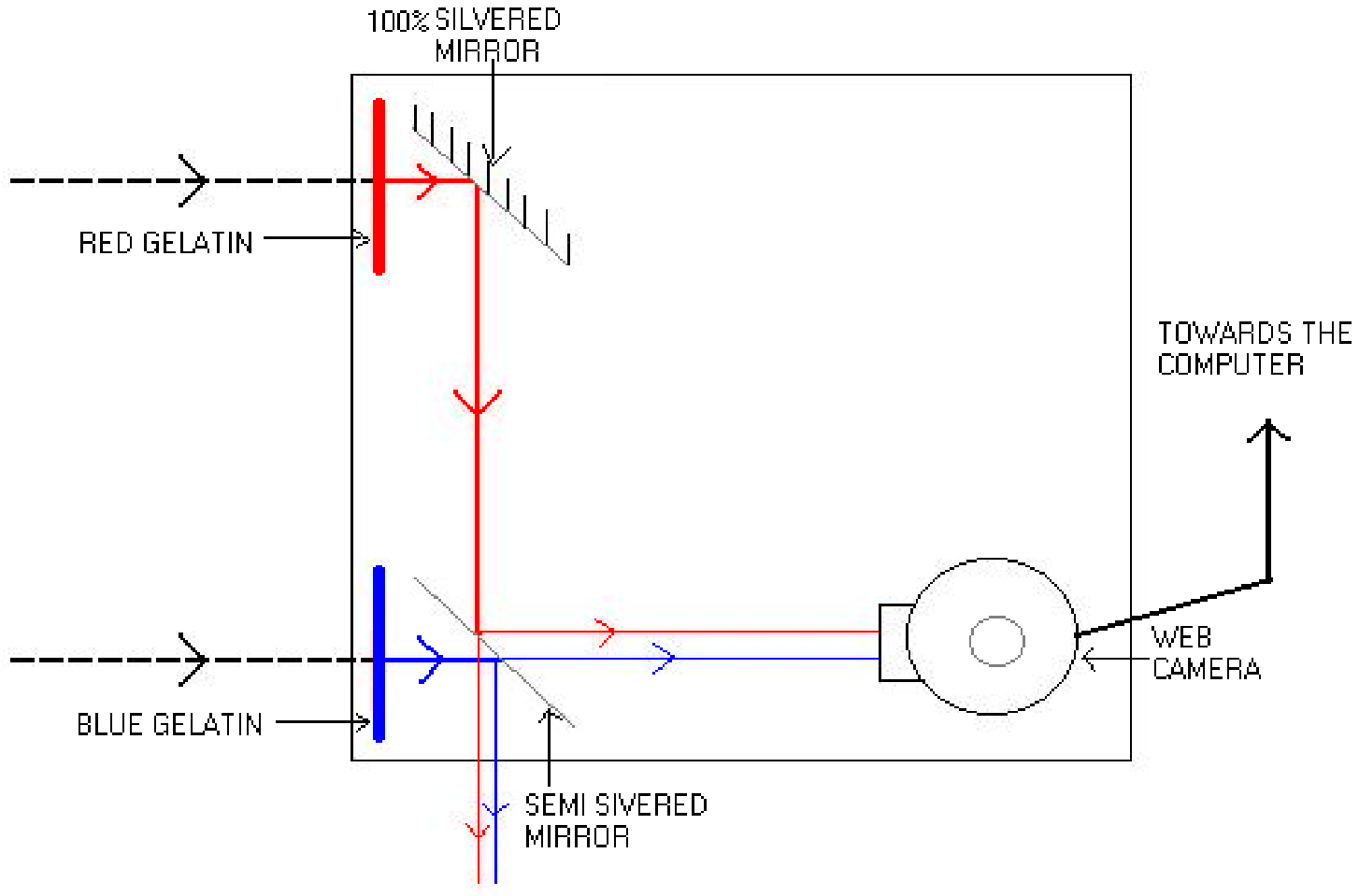}
         }
    \end{center}

\end{figure}

Figure~\ref{fig:anaglyph} shows the arrangement of the mirrors, filters and the web-camera. The anaglyph which would be obtained from the proposed model would  occupy an entire frame on the screen of the computer to which the web-camera is connected.

\clearpage
\section{Conclusion}

\indent 

The novel method of obtaining stereographs is proposed. Unlike other methods it uses only one lens. Experimental setup along with the results is also discussed in detail here. On the basis of our experiments and the success of the prototype model on stereographs, we conclude that the idea proposed here of getting stereographs using a single lens and an arrangement of mirrors, is feasible. The method can be improved by using good quality CCD camera and by reducing the overall size of the setup. Possible uses are also discussed in the paper. The proposed method of obtaining anaglyphs is also explored.

{\textheight 900pt
  \begin{center}
   {\bf Acknowledgment}
  \end{center}
}
\textheight 680pt

\indent We thank the SCI POP laboratory of Inter University Center for Astronomy and Astrophysics(IUCAA) for providing the necessary facilities.    

 One of us (KD) wishes to thank Nilesh Puntambekar for his guidance through the implementation of this idea.

        This project was originally done in a conceptual form for Kishore Vaigyanik Protsahan Yojana (KVPY) Fellowship funded by Department of Science and Technology, Government of India, under the administrative body of Indian Institute of Sciences(IISc), Bangalore and Indian Institute of Technology(IIT), Bombay. They have appreciated this innovative idea and awarded the prestigious undergraduate KVPY fellowship for the same.

\end{document}